\documentclass[aps,prd,showpacs,amssymb]{revtex4}
\usepackage{graphicx} 
\usepackage{bm}
\newcommand{\beq}{\begin{equation}}
\newcommand{\eeq}{\end{equation}}
\newcommand{\ba}{\begin{array}}
\newcommand{\ea}{\end{array}}
\newcommand{\lsim}   {\mathrel{\mathop{\kern 0pt \rlap
  {\raise.2ex\hbox{$<$}}}
  \lower.9ex\hbox{\kern-.190em $\sim$}}}
\newcommand{\gsim}   {\mathrel{\mathop{\kern 0pt \rlap
  {\raise.2ex\hbox{$>$}}}
\lower.9ex\hbox{\kern-.190em $\sim$}}}

\begin{document}
\title{Phenomenological constraints  on low-scale gravity}
\author{Veniamin Berezinsky}
\affiliation{INFN, Laboratori Nazionali del Gran Sasso, I-67010
  Assergi (AQ), Italy\\
and  Institute for Nuclear Research of the RAS, Moscow, Russia}
 \author{Mohan Narayan}
\affiliation{Mumbai University, Institute of Chemical Technology,
Mumbai 400076, India\\
and INFN, Laboratori Nazionali del Gran Sasso, I-67010, Asergi (AQ),
Italy}

\begin{abstract}
We study the constraints on gravity scale $M_P$ in extra-dimension
gravitational theory, obtained from gravity-induced processes.  
The obtained constraints are subdivided into strong (though not robust) 
and reliable (though less strong). The strong constraints can be 
in principle relaxed due to some broken gauge symmetries, e.g. 
family symmetry. The strongest constraint is given by neutrino 
oscillations. For different assumptions the lower bound on $M_P$
is $10^{15} - 10^{18}$~GeV. However, it can be, in principle, 
reduced by broken family symmetry. More reliable bounds 
are due to flavor-conserved operators or those which change the 
flavors within one family. These bounds, obtained using the electron
mass and width of $\pi \to e\nu$ decay, are $1\times 10^5$~GeV
and $5\times 10^5$~GeV, for these two cases, respectively.  
\end{abstract}
{
\maketitle

\section{\label{sec:introduction} Introduction}
Starting from works \cite{extr-dim},
there has been intense activity in the recent past in constructing
theories, in which the fundamental gravity scale $M_P$  
is much less than the usual Planck scale 
$M_{\rm Pl} \sim 1\times 10^{19}$ GeV. 
These theories necessarily involve more than four
space-time dimensions, where the extra dimensions are
compactified. Gravitation propagates in all the dimensions, while the standard
model fields are usually restricted to four dimensions or can
propagate in the extra dimensions in more complicated versions
of this scenario.
One of the main motivations of these models is to provide an 
alternative solution to
the hierarchy problem, which is usually taken care of by invoking 
supersymmetry.
The fundamental scale of gravity $M_P$ is related to
the Planck scale $M_{\rm Pl}$ via the number of extra dimensions 
involved and their
radii of compactification. 
There is an extensive literature on this subject.
We indicate recent reviews \cite{GiuWe,apl,lrrr,mpr}, in which the reader can
find references to most of the relevant papers in this area. 

Many phenomenological restrictions on the fundamental scale $M_P$ 
have been studied in the references cited above and it was demonstrated that 
low-scale gravity with $M_P \sim$~TeV survives. In this paper we shall
study restrictions imposed by quantum gravity operators, which have
been already discussed in \cite{zud}.

The general approach to description of gravity-induced processes consists in
following. 

The unknown quantum gravity Lagrangian is assumed to be expanded at
low energies in series of unrenormalizable operators, each being
inversely proportional to the powers of the fundamental scale $M_P$
\cite{planck1}:
\begin{equation}
{\cal L}(\psi,\phi) \sim \frac{{\cal O}(1)}{M_{\rm P}}(\psi\psi\phi\phi)
+\frac{{\cal O}(1)}{M^2_{\rm P}}(\psi\psi\psi\psi)+ ...~~,
\label{expans}
\end{equation}
where $\psi$ and $\phi$ are fermion and boson fields, respectively.\\
The inverse
proportionality to $M_{\rm P}$ is a natural condition of vanishing of
these operators when $M_{\rm P} \to \infty$, i.e. when gravity is
switched off. Assuming the coefficients ${\cal O}(1)$ in expansion 
(\ref{expans}) we follow argumentation of Hawking \cite{hawking}. 
The Lagrangian ${\cal L}$ and the operators of its expansion (\ref{expans})
are expected to  break the global symmetries \cite{hawking,planck1,linde}. 
It could be understood, for
example, as absorbing a global charge by virtual black hole with its
consequent evaporation. The topological fluctuations (wormhole effects)
break conservation of global charges, too \cite{wormhole}. 
It could be understood as transition of global charges to baby universes.
On the other hand the discussed
operators should respect the gauge symmetries and gauge discrete
symmetries.  In particular, the Lagrangian (\ref{expans}) must have 
SU(2)$\times$U(1) symmetry for the Standard Model (SM) fields, before this
symmetry is spontaneously broken.

The exact or  spontaneously broken gauge symmetries can in principle suppress 
some gravity operators. Berezhiani and Dvali \cite{zud} considered the 
family symmetry as mechanism of such
suppression. An example can be given by the flavor-changing
neutral currents (FCNC). If flavors are considered as global charges, 
FCNC, e.g. $\bar{d}s$, appear as gravity induced
operators being suppressed by powers of $M_P$.  FCNC effects
should be small, restricted by $K^0-\bar{K}^0$ mixing and $\mu \to e\gamma$ 
decay. Treating this problem straightforwardly, one obtains the strong
lower bound on $M_p$. As was demonstrated by Berezhiani and 
Dvali \cite{zud} the  broken
family symmetry may suppress additionally FCNC gravity operators. This
interesting possibility reduces, however, the beauty of the original 
theory: the gauge family bosons have to propagate in the bulk, what 
was prerogative of graviton only. In addition, it also seems somewhat
artificial that the gauge family bosons have this property, while the
other standard model gauge fields do not.  

There could be some other broken gauge symmetries which can suppress 
additionally the gravity operators. In this paper we shall consider 
the broken chiral symmetry as a mechanism of such suppression. 

The plan of the paper is as follows.

We begin  discussion with the family-flavor violating processes 
which in principle can be suppressed by family symmetry. Some of these 
processes have been already discussed, nevertheless we discuss them 
again, performing more
accurate calculations. Not surprisingly, for this class of processes
we obtain the strongest restrictions. Then we perform the
calculations for processes which cannot be suppressed by family
symmetry. In the section \ref{summary} we  discuss how these  
operators can be suppressed. 

\section{Processes with flavor violation }
Quantum gravity  interactions are supposed to violate the global
symmetries, and thus the flavors of particles. However, they must
respect gauge symmetries, and the broken gauge symmetries can suppress 
the flavor-changing process. In this section we
shall obtain the bounds to the  fundamental gravity scale 
$M_P$ from flavor-changing processes neglecting possible symmetry 
suppression, and keeping in mind that these bounds are not robust.  

\subsection{Constraint from $K^0-\bar{K}^0$ mixing \label{cav}}
$K^0-\bar{K}^0$ mixing is the consequence of an $\Delta S = 2$
interaction, which is very well accounted for within the framework
of the Standard Model. This mixing results in two mass eigenstates
$K_1$ and $K_2$ with masses $m_1$ and $m_2$ .

The mass difference between $K_1$ and $K_2$ is measured very precisely and
is known to be $5 \times 10^9~\mbox{sec}^{-1}$ \cite{pdg}. 
It has been realised since a
long time that such a small value of this mass difference will be very sensitive
to any new $\Delta S = 2$ contributions coming from beyond the standard model
and hence will be able to strongly constrain such new contributions.

If one has an effective $\Delta S =2$ four-fermion interaction, then 
the usual calculation for this mass difference is done using the "vacuum
dominance approximation" for the hadronic matrix element,
$\langle K^0|\bar{d}\gamma_{\mu}(1-\gamma_{5})s]
[\bar{d}\gamma_{\mu}(1-\gamma_{5})s]|\bar{K}^0 \rangle$,
(see for example \cite{okun}). This gives,
$\Delta m_{12} \approx G_2 f_{K}^2 m_K$,
where $f_K = 160~\mbox{MeV}$ \cite{pdg} is the constant appearing 
in the decay amplitude of $K^+ \rightarrow \mu^+ \nu$ and
$m_K = 494$ MeV, is the $K^+$ mass. 
$G_2$ parametrizes the strength of the new interaction and the approximation
sign can be interpreted to imply that, there could
be some dimensionless $\lambda_{\alpha \beta}$ coefficients in front of $G_2$,
where $\alpha$ and $\beta$ are the quark flavors involved in the interaction.
In our case, since low scale gravity is responsible for this contribution, the 
interaction is flavor blind and we set $\lambda_{\alpha \beta} \approx 1$.

We calculate effect of the hadronization using a scalar current in
exact analogy with $\pi$ decay (see below):
\begin{equation}
\langle 0|\bar{d} \gamma_{5}s |\bar{K}^0\rangle = f_{sK},
 \mbox{ with } f_{sK} = \frac{m^2_{K^+}}{(m_s+m_d)} f_K.
\label{deffsK}
\end{equation}

Hence we obtain
\begin{equation}
\frac{1}{M_P^2} \frac{f_{sK}^2}{m_K} < 5 \times 10^9\; \mbox{s}^{-1},
\end{equation}
resulting in
\begin{equation}
M_P > 10^7~\mbox{GeV} .
\end{equation}

\subsection{$\mu \rightarrow e \gamma$}
The lepton number violating process $\mu \rightarrow e \gamma$
gives another bound on the fundamental gravity scale  $M_P$.
The decay rate of $\mu \rightarrow e \gamma$
in the Standard Model, extended with massive neutrinos 
is given by 
$\Gamma_{\rm EW} = m_{\mu}^3 F^3/8 \pi $, where formfactor 
$F$ calculated \cite{palmoh} from the loop diagrams with exchange by 
neutrino, is very small
\begin{equation}
F=\frac{e G_F m_{\mu}}{32 \pi^2}\;L,
\label{formfactor}
\end{equation}
with $L \sim (m_{\nu}/ m_W)^2$.

The lowest order quantum gravity operator (\ref{expans}) which induces 
$\mu \rightarrow e \gamma$ decay is given by 
\begin{equation}
\mathcal{L} = \frac{1}{M_P^2} \phi \bar{e} \sigma_{\mu\nu} \mu\; 
F_{\mu\nu} ,
\label{L}
\end{equation}
where $\phi$ is the SM scalar field and $F_{\mu\nu}$  is e-m tensor
field. 

The width of $\mu \rightarrow e \gamma$ decay calculated using
operator (\ref{L}) after EW symmetry breaking is 
\begin{equation} 
\Gamma = \frac{m_{\mu}^3 v_{\rm EW}^2} {8 \pi M_P^4} ,
\label{width}
\end{equation}
where vacuum expectation value is $v_{\rm EW}=\langle \phi
\rangle = 174$~GeV \cite{pdg}. 
The branching ratio $B$ for this decay with respect to the dominant
muon decay $\mu \rightarrow e \bar{\nu_{e}} \nu_{\mu} $ , 
\begin{equation}
B = \frac{24 \pi^2 v_{\rm EW}^2}{M_P^2 G_F^2 m_{\mu}^2},
\label{branching}
\end{equation}
results in $M_P > 2\times 10^7$~GeV, if to use the present upper bound 
$B < 1.2\times 10^{-11}$ for $\mu \rightarrow e \gamma$ decay.

\subsection{Neutrino oscillations} \label{oscillations}
Here we consider the gravitational interaction among the neutrinos as 
responsible for generating an additional term in the neutrino mass matrix,
which is usually taken to be generated by some GUT dynamics.
This additional term is taken as an perturbation to the GUT mass matrix.
We follow the same approach and formalism developed 
in the work \cite{vmb} (where the main aim was to consider 
the effect of this perturbation on the neutrino mixing angles). 
We refer the reader to it for further details.
 
Because of this perturbation, the mass
splittings generated by the GUT mass matrix get an additional contribution
due to the gravitational interaction.
From Eq.(11) of \cite{vmb}, we get an expression for the modified
mass splittings
($M_{i}$ are the neutrino masses and
$\Delta M^2_{ij}=M_i^2-M_j^2$ are the mass splittings produced as a result
of some GUT texture which is able to reproduce the observed
masses and mixings). 
\begin{equation}
\Delta M^{'2}_{ij} = \Delta M^2_{ij} + 2 \mu\; (M_i \mbox{Re}[m_{ii}]
-M_j  \mbox{Re}[m_{jj}]),
\label{dms}
\end{equation}
where $m = U^{t} \lambda U$, with $U$ being the usual neutrino mixing matrix for
three flavors, $\lambda$ is a $3\times 3$ matrix with all elements
equal to $1$ and $\mu = v_{EW}^2/M_P$. 

The KamLAND \cite{kamland} measurement of the lower mass splitting
is a very robust result in neutrino physics confirming the earlier
conclusions drawn from the solar and the atmospheric neutrino data.
In the following analysis, we use this result.
KamLAND gives the central value of $\Delta M^{2}_{21}$ as
$7 \times 10^{-5}~\mbox{eV}^2$. Let us estimate the correction to
this scale coming from the second term in Eq.(\ref{dms}).
Consider the case where the neutrino spectrum
is hierarchial with 
$|M_1| \ll |M_2| \ll |M_3|$, where $M_3 \approx \sqrt{\Delta M^2_{31}}$ and
$M_2 \approx \sqrt{\Delta M^2_{21}}$.
Also, unless the phases in the mixing matrix take very special values,
the term $\mbox{Re}[m_{22}]$ has a value ${\cal O}(1)$.
Therefore the correction term can be written as
$2 \mu \sqrt{\Delta M^2_{21}}$.  
The {\em maximal} correction due to the considered term
can be parametrised with help of
$\alpha\equiv {\delta(\Delta M^2_{21})}/{\Delta M^2_{21}}$. 
For instance $\alpha=1$
implies a 100 \% correction. This can be translated into 
a bound on the scale $M_P$:
\begin{equation}
M_{P} > \frac{0.7}{\alpha} \times 10^{16}~\mbox{GeV}.
\end{equation}
The allowed values of $\alpha$ may be obtained from analysis of the 
KamLAND data, respecting the general restriction $\alpha < 1$

So the new scale must be even higher than a typical GUT scale.
For a quasi degenerate neutrino spectrum with neutrino
masses at $\approx 1$~eV, the constraint becomes stronger: 
\begin{equation}
M_{P} > 
\frac{0.9}{\alpha} \times 10^{18}~\mbox{GeV}.
\end{equation}
This result, however, comes with some caveats:
(1)~For the degenerate spectrum, for specific
choices of the Majorana phases it is 
possible to arrange for 
$\mbox{Re}[m_{ii}] \approx \mbox{Re}[m_{jj}]$ 
and hence to relax the constraint on $M_P$.
(2)~For very specific choices of the ${\cal O}(1)$ coefficients, 
it is possible to use low scale gravity as the {\em only} 
contribution to the neutrino masses \cite{pre}.

\section{Bound from proton decay}
Now we discuss the bound on the gravity scale $M_P$ coming from proton decay.
Even restricting our interest to $\Delta B=\Delta L$ 
operators, there are still many possible contributions:
$$
{\cal L}_{eff}\ni 
\frac{1}{M_P^2} \left[ \sum_{i=1}^6 c_i^S O_i^S + 
\sum_{i=1}^6 c_i^T O_i^T +
\sum_{i=7}^9 c_i^V O_i^V \right], 
\mbox{ with }c_i^{S,T,V}={\cal O}(1).
$$
In the above equation, the 6 operators  
of the scalar and of the tensorial type, $O_i^S$ and $O_i^T$
are 
$(q\vec{\sigma}q)(q\vec{\sigma}\ell)$,
$(q q)(q \ell)$,
$(u u )(d e)$,
$(u d)(u e)$, 
$(q q)(u e)$ and
$(d u)(q \ell)$.
The 3 operators, $O_i^V$ 
of the vectorial type are
$(q u)(q e)$,
$(q u)(d \ell)$ and
$(q d)(u \ell)$.
These operators were first written down explicitly in \cite{wein}.
It is simple to see that the weak hypercharge 
is conserved in all these. Regarding weak isospin, 
the first operator is built 
in the $I=1$ channel, the last 3 in $I=1/2$ and 
the remaining ones in $I=0$. Finally the color of the 
three quark fields is contracted with $\epsilon_{abc}$.
The flavor indices have been suppressed.
If we restrict our attention solely
to proton decay, i.e to $\Delta S = 0$ process, then the operators
$O_1$ and $O_3$ do not contribute. To
estimate the resulting bound on $M_P$,
let us consider as an example the channel $p\to e^+ \pi^0$,
that has a strong bound of $\tau=1.6\cdot 10^{33}$~years 
from the Super-Kamiokande experiment.
The typical hadronic matrix element e.g.,  
$\langle \pi^0|\epsilon_{abc}(q^a q^b) q^c_A |p\rangle$ can be
presented as $\sqrt{m_p}\; \psi(0)\; u_A$, where $\psi$ is an quark overlap 
function with value at origin $|\psi(0)|^2=\mbox{few}\cdot m_\pi^3$ and 
$u_A$ is the spinor describing the proton. 
In this way, we obtain the bound:
\begin{equation}
M_P\ge \left(
\frac{m_p\cdot \tau}{32\pi}\cdot m_p |\psi(0)|^2 \xi^2\
\right)^{1/4}\ge 5\times 10^{15}~\mbox{GeV}, 
\end{equation}
where $\xi$ is a renormalization factor from $M_P$ 
to the scale of proton decay. 
This bound is expected to
be correct within a factor of the order of unity, however it is 
remarkably insensitive to the details of the 
hadronization.
\section{Bound from the Majorana neutrino mass}
\label{neutrino-mass}
The quantum gravity dimension 5 operator can be constructed as SU(2) 
singlet  from lepton and Higgs fields: 
\begin{equation}
{\cal L } = \frac{1}{M_P}\left(l_L^t \vec{\tau} l_L^c \right) 
\left(\phi^t \vec{\tau}\phi^c \right) ,
\label{Mmass}
\end{equation}
where $l_L$ is lepton SU(2) doublet, index $c$ denotes the charge
conjugation and $\vec{\tau}$ the Pauli spin matrices.
This operator generates the Majorana neutrino mass and as it was first
suggested in Ref.~\cite{ABS}.
After EW symmetry breaking, Eq.(\ref{Mmass}) gives the following
Majorana neutrino mass term:
\begin{equation}
{\cal L_{\rm mass}} = \frac{v_{EW}^2}{M_P} \nu_{L}\nu_{L}^c.
\label{numass}
\end{equation}

From the cosmological limit on the neutrino masses, 
$m_{\nu} < 0.23$ eV \cite{wmap,disc},
valid for any flavor, we obtain using Eq.(\ref{numass}):
\begin{equation}
M_P > \frac{v_{EW}^2}{m_{\nu}} = 1.3\times 10^{14}~\mbox{GeV}.
\label{Mbound} 
\end{equation}

Note that though Eqs.~(\ref{Mmass}) and (\ref{numass}) violate lepton
number, they do not allow transitions between flavors (lepton numbers) 
of different families, and therefore the bound (\ref{Mbound}) 
cannot be suppressed by family symmetry. For further discussion 
see section \ref{summary}.

\section{Processes without family-flavor violation}
In this section we shall study more reliable bounds on the 
gravity scale $M_P$. We restrict our consideration by flavor transitions 
within one family and by flavor-conserving 
operators.  
\subsection{Fermion masses} \label{Ferm-masses}

The Standard Model includes the Yukawa interaction
\begin{equation}
{\cal L}=h(\bar{f}_L\phi)f_R,
\label{Yukawa}
\end{equation}
where $f_L$ is fermion $SU(2)_L$ doublet, $\phi$ is the Higgs doublet and 
$f_R$ is  $SU(2)_L$ singlet. The corresponding gravitational operator
can be written, including $SU(2)_L$ singlet $(\phi \phi)$:
\begin{equation}
{\cal L}_{\rm grav}= \frac{1}{M_P^2}(\bar{f}_L \phi)f_R (\phi \phi).
\label{mfer}
\end{equation}
After EW symmetry breaking, the Eq.(\ref{mfer}) gives the fermion mass term 
\begin{equation}
m_f=\frac{v_{\rm EW}^3}{M_P^2}
\label{m_f}
\end{equation}
Using  $m_f < m_e$, we obtain  the lower bound 
\begin{equation}
M_P> 1\times 10^5~{\rm GeV}
\label{fermion}
\end{equation}
However, the Higgs sector of the Standard Model is most questionable
part of the model. In particular, the differences of fermion masses 
in the first and third families requires the physics beyond SM. 

The Lagrangian (\ref{mfer}) may be forbidden 
imposing the chiral symmetry, and the lower bound on $M_P$ may
expected to become smaller. We shall demonstrate here that this is not
the case.   

Consider the chiral 
gauge symmetry within most natural  $SU(2)_L\times SU(2)_R$ 
model \cite{LR-symm}. In this model for the first family 
$q_L=(u,~d)_L$, $l_L=(\nu,~e)_L$ and $\phi_L=(\phi^+,~\phi^0)_L$ 
are transforming as $(2,~1)$ and $q_R=(u,~d)_R$,
$l_R=(\nu,~e)_R$ and $\phi'_R=(\phi^{'+},~\phi^{'0})_R$  
as $(1,~2)$. For the other families $q$ and $l$ are defined in the
identical way.  The operator (\ref{mfer}) is not 
$SU(2)_L\times SU(2)_R$ singlet, and it does not conserve chiral 
charges. One
can see it explicitly introducing the gauge interactions in the usual way:
\begin{equation}
{\cal L}=
g_L(\bar{q}_L\gamma_{\mu}\vec{\tau}q_L+\bar{l}_L\gamma_{\mu}\vec{\tau}l_L+
\phi^*_L\partial_{\mu}\vec{\tau}\phi_L)\vec{W}_{\mu L}+
g_R(\bar{q}_R\gamma_{\mu}\vec{\tau}q_R+\bar{l}_R\gamma_{\mu}\vec{\tau}l_R+
\phi^*_R\partial_{\mu}\vec{\tau}\phi_R)\vec{W}_{\mu R}
\label{L-R}
\end{equation}
Now we can write the gravity operator as  $SU(2)_L\times SU(2)_R$ singlet
which conserves the chiral charges $g_L$
and $g_R$:
\begin{equation}
{\cal L}_{\rm singl}= \frac{1}{M_P}(\bar{f}_L \phi_L)(f_R \phi_R).
\label{singl}
\end{equation}
After spontaneous symmetry breaking we obtain for the fermion mass 
\begin{equation}
m_f=\frac{1}{M_P}v_{\rm EW}v_R,
\label{m_e}
\end{equation}
which should be less than $m_e$. The value of $v_R$ is unknown, but it
cannot be less than $v_{\rm EW}$. 
Then one obtains
\begin{equation}
M_P > v_{\rm EW}^2/m_e=5.9\times 10^7~{\rm GeV}.
\label{bound-emass}
\end{equation}
Thus, submerging SU(2)$_{\rm L}\times U(1)$ into SU(2)$_{\rm L}\times
{\rm SU(2)_R}$ which respects chiral symmetry, we further increase the 
lower bound on $M_P$. Further on we shall use only 
limit (\ref{fermion}) as being more conservative.

\subsection{Bound from $\pi \to e\nu$}
Let us consider the operator
\begin{equation}
{\cal L}=\frac{1}{M_P^2}(\bar{l}_L\gamma_{\mu}q_L)
(\bar{q}_R\gamma_{\mu}l_R),
\label{beta}
\end{equation}
where $l_L$ and $q_L$ are $SU(2)_L$ doublets and $l_R$ and $q_R$ are 
$SU(2)_L$ singlets. The operator (\ref{beta}) 
is allowed also by $U(1)_L\times U(1)_R$ chiral symmetry, if
this symmetry is universal for all right/left  fermions.  
See section \ref{summary} for further discussion of chiral-symmetry 
restrictions. 

Therefore, all known symmetries and 
conservation laws are satisfied with (\ref{beta}), 
and we apply it to one family to avoid possible 
family-symmetry restrictions.
In terms of the first generation fields it gives
\begin{equation}
{\cal L}=\frac{1}{M_P^2}(\bar{\nu}_L\gamma_{\mu}u_L)
(\bar{d}_R\gamma_{\mu}e_R).
\label{beta1}
\end{equation}

Performing the Fierz transformation \cite{okun} for operator (\ref{beta1})
 we obtain 
\begin{equation}
{\cal L} = \frac{2}{M_P^2} (\bar{u}(1+\gamma_5)d)
(\bar{e}(1-\gamma_5)\nu_e) .
\label{beta3}
\end{equation}
This operator results in two observational consequences: (i) it gives 
the matrix element for $\pi \to e\nu$-decay unsuppressed by factor 
$m_e/m_{\pi}$, as it occurs in weak interaction, and (ii) it produces 
right-handed electrons in beta decays with the opposite helicity in 
comparison with weak interaction. 

We shall concentrate here on much stronger effect (i). 

For practical calculations it is enough to use the operator 
\begin{equation}
{\cal L_{\rm eff}} = \frac{2}{M^2_P}
(\bar{u}\gamma_{5} d)(\bar{e}(1-\gamma_{5})\nu_e),
\label{neupi}
\end{equation}
where we omitted the scalar contribution in the $\bar{u}d$
current since it vanishes between hadronic states  $<0|$ and $|\pi>$.
The standard calculations can be performed using the matrix element 
for pion decay $\langle 0|\bar{u}\gamma_{\mu}\gamma_5 d |\pi\rangle
=f_{\pi}p_{\mu}$, where $p_{\mu}$ is pion momentum and 
$f_{\pi}=130$~MeV \cite{pdg}. Multiplying this equation to $p_{\mu}$ and using
the Dirac equations for $u$ and $d$ quark fields, one obtains 
\begin{equation}
\langle 0|\bar{u}\gamma_{5}d(0)|\pi\rangle = 
 \frac{m^2_{\pi}}{m_u+m_d} f_{\pi} ,
\label{deffsx}
\end{equation}
where the quark masses $m_{u}$ and $m_{d}$ are taken to
be 4.5 and 8.5~MeV, respectively \cite{pdg}. 

Using Eq.(\ref{deffsx}), we evaluate 
the width of $\pi_{e2}$ decay, from the contribution coming from weak
interaction and and from Eq.(\ref{neupi}): 
\begin{equation}
\Gamma_{\rm tot} = \frac{G_F^2}{8 \pi}f^2_{\pi} m_{\pi} m^2_{e}
\left[1 -\frac{1}{M^2_P}
\frac{m^2_{\pi}}{(m_u+m_d)} \frac{2\sqrt{2}}{G_{F}m_e}\right]^2.
\label{newwidth}
\end{equation}
However, the Standard Model prediction for the decay width $\Gamma$ agrees  
with the experimental width within the error 
$R = \delta\Gamma_{X}/\Gamma
=4 \times 10^{-3}$ \cite{pdg}, and this limits 
the new contribution. Thus:
\begin{equation}
M_{P} > \left[\frac{1}{R}
\frac{m^2_{\pi}}{(m_u+m_d)} \frac{2 \sqrt{2}}{G_{F}m_e}\right]^{1/2}
\approx 5 \times 10^5~\mbox{GeV},                
\end{equation}

\section{Summary and discussion} \label{summary}

The status of the restrictions obtained for scale $M_P$ varies when different
processes are used.

The constraints from $K^0 - \bar{K}^0$ transition and 
$\mu \to e\gamma$ decay can be straightforwardly suppressed by the 
Berezhiani-Dvali mechanism \cite{zud}. Gravity-induced neutrino
oscillations in principle is also suppressed by family symmetry,
but this suppression would appear as well in the standard mechanism
for neutrino oscillations, destroying its agreement with observational
data. This problem can be probably solved in a model-dependent way 
(see section \ref{oscillations}). In particular, $M_P$ bound decreases when
neutrino masses are produced due to gravitational effects only. 

The bound due to Majorana neutrino mass is more robust than one due to
oscillations. It is not suppressed by family symmetry and occurs if
quantum gravity violates lepton numbers. 
The case of (almost) Dirac neutrino is given by smallness of this
term, which implies even stronger bound on $M_P$.   A possible
relaxation of this bound can be imposed in theory where lepton number
is a gauge charge. Violation of lepton number is provided by breaking
of this symmetry, and then operator (\ref{Mmass}) can include
additional small factor, which reduces the bound (\ref{Mbound}).
Notice, however, that if such mechanism allows transition between
lepton numbers in different families, it affects the standard
oscillation models, like in the case discussed above.

Since the proton decay is not discovered, the existence of some symmetry
protecting the proton and 
respected by gravitational interaction cannot be excluded (see 
Antoniadis et al. in Ref. \cite{extr-dim}). 
Note that the family symmetry \cite{zud} 
does not work in this case. Moreover, it is always possible to find
some ad hoc mechanism of suppression which makes proton ``practically
stable''(e.g. see \cite{BeSm}). Using all these arguments one may probably
exclude bound due to proton decay from the list of robust restrictions.

The bounds from fermion masses and $\pi \to e\nu$ are practically  robust. 
The former does not include even global symmetry violation and 
flavor-changing currents. The operator (\ref{singl}) is one-family
operator, it respects chiral symmetry $SU(2)_L\times SU(2)_R$, being a
singlet of this group. 
If no $SU(2)_R$ sector exists, and the symmetry is reduced to 
$SU(2)_L\times U(1)$ of the SM, the electron mass is given by 
operator (\ref{mfer}). 

The operator (\ref{beta}) for $\pi \to e\nu$ is one-family operator, 
conserving
hypercharge, lepton and baryon numbers. Being $SU(2)_L\times SU(2)_R$
singlet it respects the chiral symmetry. This decay is described by 
the flavor-changing current allowed in CC weak interaction,
and it must be allowed in quantum gravity. No global symmetry is violated
by this operator.

Therefore, all theoretically known symmetries allow the operators 
(\ref{singl}) and (\ref{beta}). In principle, these operators can be
suppressed by a new chiral symmetry, given for example by the
following chiral transformations:
\begin{equation}
l_L \to e^{i\theta_L^l}l_L,\;\; q_L \to e^{i\theta_L^q}q_L,\;\;
f_R \to e^{i\theta_R}f_R,
\label{AHCS}
\end{equation}
where the rotation angles $\theta_L^l$,~  $\theta_L^q$, and 
$\theta_R$ are not equal. It results in the different
coupling constants $g_L^l$,~ $g_L^q$ and $g_R$ in interaction of
fermions with gauge bosons $A_{L,l}^{\mu}$,~ $A_{L,q}^{\mu}$,
and $A_R^{\mu}$. 

Operator (\ref{beta}) is explicitly forbidden by this ``ad hoc chiral 
symmetry'' (AHCS) , and the modified AHCS-conserving operators are
suppressed when this symmetry is broken. Operator (\ref{singl}) can be
suppressed also, but transformation of the Higgs fields must be
specified. 

This particular chiral symmetry meets two problems: 
the model is not anomaly-free because $g_L^q \not= g_L^l$, and for the
simplest Higgs sector the EW symmetry breaking occurs at the AHCS scale.
Indeed, for the massless fields one should introduce two scalars 
$\phi_l$ and $\phi_q$ to provide masses for $A_{L,l}$ and $A_{L,q}$
after spontaneous symmetry breaking. However, vev's 
$\langle \phi_l\rangle$ and $\langle \phi_q\rangle$ break also EW symmetry
giving too large masses to W- and Z-bosons. 

Finally, AHCS symmetry does not fit the GUT models, while 
$SU(2)_L\times SU(2)_R$ symmetry is typical for most GUT models, most 
notably for SO(10).

Probably one can construct more complicated AHCS model
with the problems indicated above being  solved. However, in absence of any
motivation for symmetry (\ref{AHCS}) we qualify this possibility as
exotic.

Obtained bounds are summarised in Table \ref{tab1}.

\begin{table}[h]
\caption{\em Bounds on the gravity scale $M_P$ from
various processes.} 
\begin{center}
\begin{tabular}{|l||c|}
\hline\hline
Process&Lower bound on $M_P$ \\ \hline\hline
{\em Within single family:} \\ \hline\hline
Electron mass & $10^5$ GeV \\ \hline
$\pi_{e2}$ decay &  $5 \times 10^{5}$ GeV \\ \hline
Majorana neutrino mass & $10^{14}$ GeV \\ \hline
Proton decay &  $10^{15}$ GeV  \\ \hline\hline
{\em Transitions between families:} \\ \hline\hline
$K^0-\bar{K}^{0}$ oscillations &  $1 \times 10^{7}$ GeV \\ \hline
$\mu \rightarrow e \gamma$ &  $1 \times 10^{7}$ GeV \\ \hline
Neutrino oscillations & $10^{15}-10^{18}$ GeV \\ \hline\hline
\end{tabular}
\end{center}
\label{tab1}
\end{table}

In conclusion, the supergravity operators (\ref{expans}) impose the
lower limits on the fundamental gravity scale $M_P \gg$~TeV. These
limits are valid even if supergravity operators (\ref{expans}) break only
fermion flavors, like in case of $\pi \to e\nu$ decay, or conserves
all flavors like in case of a fermion mass.  The obtained lower limits
must be considered as a problem for TeV-scale gravity, though we do
not interpret these results as exclusion of TeV-scale gravity. The
simplest possibility is given by some symmetry which forbids or
suppresses the dangerous operators like the AHCS considered
above. However, TeV-scale gravity is only a possibility,  
and the lower limits 
obtained above, could be an argument in favor of extra-dimension 
gravitational theory with the larger scale, like e.g.   $M_P \sim 10^{15}$~GeV 
in the Horava-Witten scenario \cite{HW}. These models have the interesting
phenomenological applications, like for example the gravity-induced 
neutrino masses ( see section \ref{neutrino-mass} and \cite{pre}).

\section*{Acknowledgements}
We are grateful to Francesco Vissani for the collaboration and  
detailed discussions of all items of this paper. 
We thank Valery Rubakov, Goran Senjanovic, Alex Vilenkin,
and Valentine Zakharov for valuable discussions. The anonymous 
referee is thanked for useful remarks and criticism.


\begin{thebibliography}{99}

\bibitem{extr-dim}
N.~Arkani-Hamed, S.~Dimopoulos, G.~Dvali, Phys. Lett. {\bf B 429}, 263 
(1998),\\
I.~Antoniadis, N.~Arkani-Hamed, S.~Dimopoulos, G.~Dvali, 
Phys. Lett. {\bf B 436} 257 (1998),\\
N.~Arkani-Hamed, S.~Dimopoulos, G.~Dvali,  Phys. Rev. {\bf D 59}, 086004 
(1999).

\bibitem{GiuWe}
G.~F.~Giudice and J.~D.~Wells, Review of Particle Physics (Particle Data 
Group) J. Phys. G {\bf 33}, 1 (2006).

\bibitem{apl}
A.~Perez-Lorezana, AIP~Conf.~Proc {\bf 562}, 63 (2001),
(hep-ph/0008333).


\bibitem{lrrr}
A.~Lukas, P.~Ramond, A.~Romanino and G.~G.~Ross,
JHEP {\bf 0104}, 010 (2001).

\bibitem{mpr}
A.~Muck, A.~Pilaftsis and R.~Ruckl, hep-ph/0209371 
(Lectures given by R.~Ruckl at the international school,
"Heavy Quark Physics", JINR, Dubna).

\bibitem{zud}
Z.~Berezhiani and G.~Dvali,
Phys. Lett. B {\bf 450}, 24 (1999).


\bibitem{planck1}
R.~Barbieri, J.R.~Ellis and M.K.~Gaillard,
Phys. Lett. B {\bf 90}, 249 (1980).

\bibitem{hawking}
S.~W.~Hawking,
Phys.\ Lett.\ A {\bf 60}, 81 (1977),\\
S.~W.~Hawking, D.~N.~Page, C.~N.~Pope, Phys. Lett. {\bf 86}, 175 (1979),\\
S.~W.~Hawking, D.~N.~Page, C.~N.~Pope, Nucl. Phys. {\bf B 170}, 283 (1980).


\bibitem{linde}
R.~Kallosh, A.~Linde, D.~Linde, L.~Susskind, Phys. Rev. {\bf D 52}, 912 (1995).

\bibitem{wormhole}
G.V. Lavrelashvilli, V. A. Rubakov and P. G. Tinyakov, Nucl. Phys. B
{\bf 299}, 757 (1988).
S. Coleman, Nucl. Phys. B {\bf 307}, 867 (1988).
S. B. Giddings and A. Strominger, Nucl. Phys. B {\bf 307}, 854 (1988).

\bibitem{pdg}
"The Review of Particle Properties",
K.~Hagiwara {\it et al}, PRD {\bf 66}, 010001 (2002).

\bibitem{okun}
L.~B.~Okun, Leptons and Quarks
(North Holland Publishing House) (1987).

\bibitem{palmoh}
See for example, "Massive Neutrinos in Physics and Astrophysics", by
P.~B.~Pal and R.~N.~Mohapatra
(World Scientific Publishing) (1998).

\bibitem{vmb}
F.~Vissani, M.~Narayan and V.~Berezinsky,
Phys. Lett. B {\bf 571}, 209 (2003).

\bibitem{kamland}
K. Eguchi et al. (KamLAND collaboration)
Phys. Rev. Lett {\bf 90}, 021802 (2003).

\bibitem{pre}
V.~Berezinsky, M.~Narayan and F.~Vissani, JHEP {\bf 0504}, 009 (2005). 

\bibitem{wein}
S. Weinberg, Phys. Rev. Lett. {\bf 43}, 1566 (1979) .

\bibitem{ABS}
E.~Akhmedov, Z.~Berezhiani, G.~Senjanovic, Phys. Rev. Lett. {\bf 69},
3013 (1992) .

\bibitem{wmap}
D.N.~Spergel {\it et al.},
astro-ph/0302209.
Other relevant data and 
analyses are in \O.~Elgaroy {\it et al.},
Phys.\ Rev.\ Lett.\  {\bf 89}, 061301 (2002) and
A.~Lewis and S.~Bridle,
Phys.\ Rev.\ D {\bf 66}, 103511 (2002).

\bibitem{disc}
\O.~Elgaroy and O.~Lahav,
JCAP {\bf 04}, 004 (2003);
S.L.~Bridle, O.~Lahav, J.P.~Ostriker and P.J.~Steinhardt,
Science {\bf 299}, 1532 (2003).

\bibitem{LR-symm}
J.C.~Pati and A.~Salam, Phys. Rev. D {\bf 10}, 275 (1974),\\
R.N.~Mohapatra and J.C.~Pati, Phys. Rev. D {\bf 11}, 566 (1975),\\
G.~Senjanovic and R.N.~Mohapatra, Phys. Rev. D {\bf 12}, 1502 (1975) .

\bibitem{BeSm}
V.Berezinsky and A. Smirnov, Phys. Lett. {\bf B97}, 371 (1980), \\
Z.~Berezhiani, T.~Gogoladze and A.~Kobakhidze, Phys. Lett. B {\bf 522},
107 (2001). 

\bibitem{HW}
P.~Horava and E.~Witten, Nucl. Phys. B {\bf 460}, 506 (1996) and 
{\bf 475}, 94 (1996).


\end{thebibliography}
\end{document}